\documentclass[aps,prl,twocolumn,superscriptaddress,floatfix]{revtex4}
\pdfoutput=1
\usepackage{graphicx}
\usepackage{dcolumn}
\usepackage{bm}
\usepackage{amsfonts}
\usepackage{amsmath}
\usepackage[colorlinks,citecolor=blue,linkcolor=black,urlcolor=red]{hyperref}
\newcommand{\SM}{\hyperref[SM]{SM\ref{SM}}}

\begin{document}

\title{Two-dimensional Mott-Hubbard electrons in an artificial honeycomb lattice}
\author{A. Singha}
\thanks{Present address: Department of Physics, Bose Institute, 93/1 Acharya Prafulla Chandra Road, Kolkata 700009, India.}
\affiliation{NEST, Istituto Nanoscienze-CNR and Scuola Normale Superiore, I-56126 Pisa, Italy}
\author{M. Gibertini}
\affiliation{NEST, Istituto Nanoscienze-CNR and Scuola Normale Superiore, I-56126 Pisa, Italy}
\author{B. Karmakar}
\affiliation{NEST, Istituto Nanoscienze-CNR and Scuola Normale Superiore, I-56126 Pisa, Italy}
\author{S. Yuan}
\affiliation{Radboud University Nijmegen, Institute for Molecules and Materials, NL-6525 AJ Nijmegen, The Netherlands}
\author{M. Polini}
\email{m.polini@sns.it}
\affiliation{NEST, Istituto Nanoscienze-CNR and Scuola Normale Superiore, I-56126 Pisa, Italy}
\affiliation{Kavli Institute for Theoretical Physics China, CAS, Beijing 100190,  China}
\author{G. Vignale}
\affiliation{Department of Physics and Astronomy, University of Missouri, Columbia, Missouri 65211, USA}
\affiliation{Kavli Institute for Theoretical Physics China, CAS, Beijing 100190, China}
\author{M.I. Katsnelson}
\affiliation{Radboud University Nijmegen, Institute for Molecules and Materials, NL-6525 AJ Nijmegen, The Netherlands}
\author{A. Pinczuk}
\affiliation{Department of Applied Physics and Applied Mathematics and Department of Physics, Columbia University New York, USA}
\author{L.N. Pfeiffer}
\affiliation{Department of Electrical Engineering, Princeton University, Princeton, NJ, USA}
\author{K.W. West}
\affiliation{Department of Electrical Engineering, Princeton University, Princeton, NJ, USA}
\author{V. Pellegrini}
\email{vp@sns.it}
\affiliation{NEST, Istituto Nanoscienze-CNR and Scuola Normale Superiore, I-56126 Pisa, Italy}

\begin{abstract}
Electrons in artificial lattices enable explorations of the impact of repulsive Coulomb interactions in a tunable system. 
We have trapped two-dimensional electrons belonging to a gallium arsenide quantum well in a nanofabricated lattice with honeycomb geometry.
We probe the excitation spectrum in a magnetic field identifying novel collective modes that emerge 
from the Coulomb interaction in the artificial lattice as predicted by the Mott-Hubbard model.
These observations allow us to determine the Hubbard gap and suggest the existence of a novel Coulomb-driven ground state. 
This approach offers new venues for the study of quantum phenomena in a controllable solid-state system.
\end{abstract}

\maketitle

When electrons roam in a solid they experience the crystalline potential created by a periodic arrangement of coupled quantum units (such as ions, atoms, or molecules). Numerous properties of solids can be explained within the paradigmatic theory of Bloch bands~\cite{Ashcroft_and_Mermin}, which neglects Coulomb interactions. These, however, often lead to profound qualitative changes that are particularly pronounced in solids with narrow energy bands~\cite{kotliar_pt_2004,kotliar_rmp_2006,vollhardt_condmat_2010}. These {\it strongly correlated} materials display exotic ordering phenomena and metal-insulator phase transitions~\cite{kotliar_rmp_2006,vollhardt_condmat_2010}. 

Mott showed~\cite{mott_prsla_1949} that interaction-induced insulators are better described in real space (rather than in momentum space), in which the solid is viewed as a collection of localized electrons bound to atoms with partially filled shells. Electrons hopping through the lattice are absorbed and emitted from the atoms thus originating two bands, which are split by the energy cost of having two electrons with antiparallel spin on the same atomic site.

Hubbard subsequently introduced a model Hamiltonian with on-site interactions that displays split bands ({\it Hubbard bands} or HBs) in the strongly correlated (or atomic) limit~\cite{hubbard_prsla_1964}. HBs (and their coexistence with quasiparticle bands in the correlated metallic phase) are ``quintessential" features of such strongly correlated systems. 
Fermions on a honeycomb lattice, in particular, have been predicted to display unusual correlated phases of matter such as topological Mott insulating~\cite{raghu_prl_2008} and quantum spin liquid phases~\cite{meng_nature_2010}. Thus the creation of artificial systems with a high degree of tunability that offer access to the Mott-Hubbard physics is an extremely appealing endeavor~\cite{buluta_science_2009}.

Here we report on the creation of an artificial lattice with honeycomb geometry for electrons, and we demonstrate the formation of HBs due to strong correlations. The artificial lattice is obtained by nanofabrication applied to a GaAs heterostructure that supports a high quality two-dimensional electron gas (2DEG)~\cite{gibertini_prbr_2009,DeSimoni2010,park_nanoletters_2009}. We probe the excitation spectrum of the electrons by inelastic light scattering and observe signatures stemming from strong Coulomb interactions, which are tuned by the application of an external magnetic field.

We find that carriers in the patterned structures support a novel {\it collective} mode that has energy that scales like $\sqrt{B}$, where $B$ is the component of the magnetic field perpendicular to the 2DEG. A theoretical analysis based on a minimal Hubbard model reveals that the mode energy is determined by the on-site Coulomb interaction and thus represents direct evidence of the existence of HBs in the 2DEG subjected to the artificial lattice. 

At low temperatures we find evidence for the opening of an unexpected gap in the spin excitation spectrum at large $B$ fields. 
This observation underpins the impact of collective phenomena in artificial lattices.
We argue that the observed gap reveals the occurrence of a new correlated phase of electrons in a honeycomb lattice akin to one of those discussed in the context of graphene at high magnetic fields~\cite{zhang_prl_2006,checkelsky_prl_2008,giesbers_prb_2009}. These findings pave the way for the possibility to explore graphene-like physics in the ultra-high-magnetic-field limit in which the magnetic length is smaller than the lattice constant of the artificial crystal - a regime not accessible in graphene.

The capability of observing Mott-Hubbard physics in nanostructured semiconductor devices with honeycomb geometry may open new approaches for the investigation of quantum phases of strongly correlated condensed-matter systems. Given that the interaction strengths governing the physics of the 2DEG can be finely tuned by design and the application of external electric and magnetic fields~\cite{byrnes_prl_2007,byrnes_prb_2008}, such scalable solid-state systems offer great promise to further expand the current realms of study offered by quantum emulators that have been so far realized with cold atom gases in optical lattices~\cite{greiner_nature_2002,lewenstein_advancesphysics_2007,bloch_rmp_2008}.

\begin{figure*}[!ht]
\centering
\includegraphics[width=0.90\linewidth]{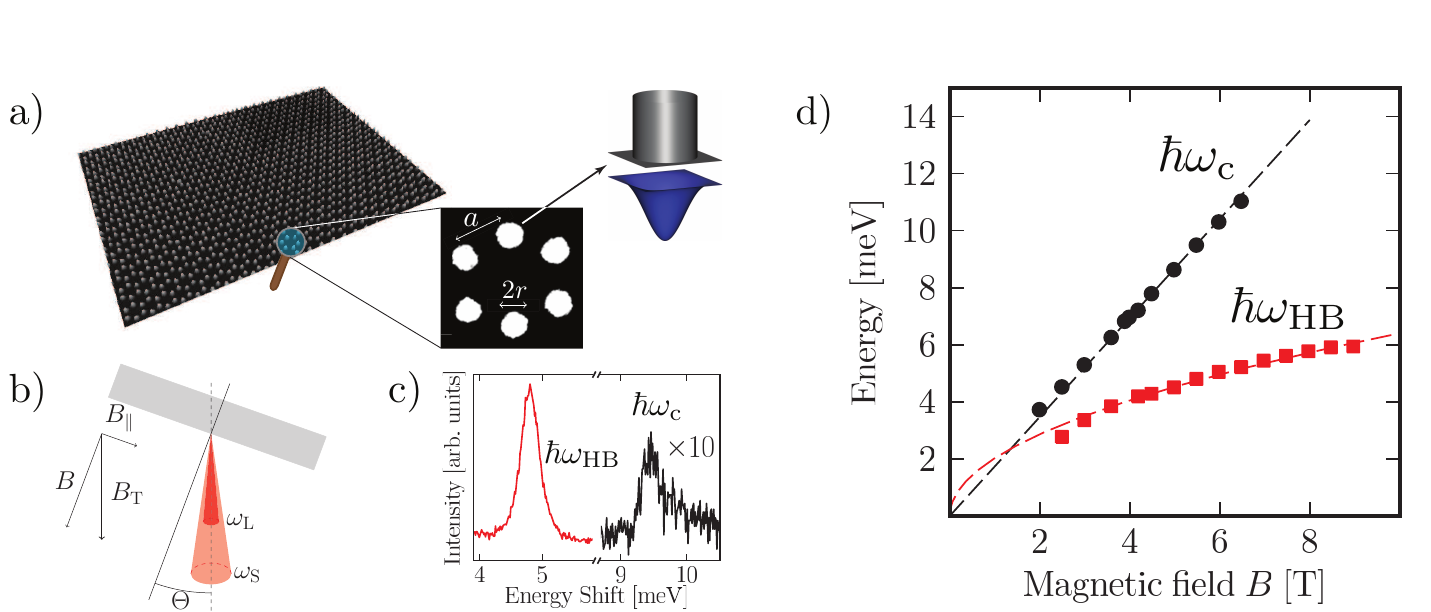}
\caption{
a) Scanning electron microscopy (SEM) image of the semiconductor artificial lattice. An expanded view of the SEM image showing a single honeycomb cell ($2 r \sim 60~{\rm nm}$, $a\sim 130~{\rm nm}$). The two-dimensional electron gas is positioned $170~{\rm nm}$ below the surface with a low-temperature mobility of $2.7 \times 10^6~{\rm cm}^{2}/({\rm V s})$.  We also sketch a cartoon of the two-dimensional potential trap for electrons induced by the nanofabricated pillar at the surface. b) Geometry of the light scattering experiment: $\omega_{\rm L,S}$ labels the incident (scattered) photon energy and $\Theta = 5^\circ$ is the tilt angle. c) Resonant inelastic light scattering spectra showing the cyclotron mode and the new low-lying collective mode at $B = 5.48~{\rm T}$ and $T = 1.7~{\rm K}$. d) Evolution of the energies of the cyclotron mode (black filled circles) and of the new collective mode at frequencies $\omega_{\rm HB}$ (red filled squares) at $T = 1.7~{\rm K}$. The  black dashed line is a linear fit to the data using $\hbar \omega_{\rm c} = \hbar eB/(m^{*}c)$. We find $m^{*} = 0.067~m_{\rm e}$ with $m_{\rm e}$ the bare electron mass, in agreement with the  bulk GaAs value. The red dashed line is a fit with $\hbar\omega_{\rm HB} = \alpha \sqrt{B[{\rm T}]}$ and $\alpha \sim 2~{\rm meV}$.\label{fig:one}}
\end{figure*}

The sample used in this study is the host of a 2DEG in a $25~{\rm nm}$ wide, one-side modulation-doped  Al$_{0.1}$Ga$_{0.9}$As/GaAs  quantum well. The procedures for nanofabricating the artificial lattice ~\cite{garcia,gibertini_prbr_2009} are detailed in the Supplementary Material (\SM). The artificial honeycomb lattice extends over a $100~{\rm \mu m}\times 100~{\rm \mu m}$ square region with a lattice constant $a \sim 130~{\rm nm}$ [see Fig.~\ref{fig:one}a)]. We denote by $V_0$ the amplitude of the artificial lattice potential. Here we focus on a sample with an estimated~\cite{DeSimoni2010} $V_0 \sim 4~{\rm meV}$ and electron density after processing $n_{\rm e} \sim 3-4 \times 10^{10}~{\rm cm}^{-2}$ corresponding to an average number of eight electrons per site. 

The inelastic light scattering experiments were performed in a backscattering configuration [see Fig.~\ref{fig:one}b)] in the temperature
range $50~{\rm mK} - 4~{\rm K}$. The light scattering technique gives direct access to the collective modes of the system that manifest as sharp peaks in the intensity of the scattered light at a given energy shift from the laser energy. Crucial for these observations is the resonant
enhancement of the light scattering cross section that occurs as the incident laser energy is scanned across an inter-band transition of the host GaAs semiconductor (see \SM).

The nanostructured 2DEGs displays well-resolved quantum Hall signatures below $3~{\rm T}$ with the honeycomb potential manifesting itself in a modulation of the magneto-resistivity periodic in $B$ (data not shown and Ref.~\onlinecite{DeSimoni2010}). At larger fields, an increase in the longitudinal resistivity signals a crossover to a regime of suppressed inter-site hopping in which new collective modes emerge. 

In addition to the ordinary cyclotron mode [black filled circles in Fig.~\ref{fig:one}d)] at energy $\hbar \omega_{\rm c} = \hbar eB/(m^{*}c)$, $m^{*}$ being the GaAs electron effective mass~\cite{pinczuk92}, the light scattering spectra display a new mode at lower energies as shown in Figs.~\ref{fig:one}c) and~\ref{fig:one}d).  This new excitation reveals its collective character in the sharpness and intensity of the light scattering peak~\cite{sokratis}. The surprising sublinear dependence of the energy of the new mode on $B$ is shown in Fig.~\ref{fig:one}d) (red filled squares).

\begin{figure}[!ht]
\centering
\includegraphics[width=0.90\linewidth]{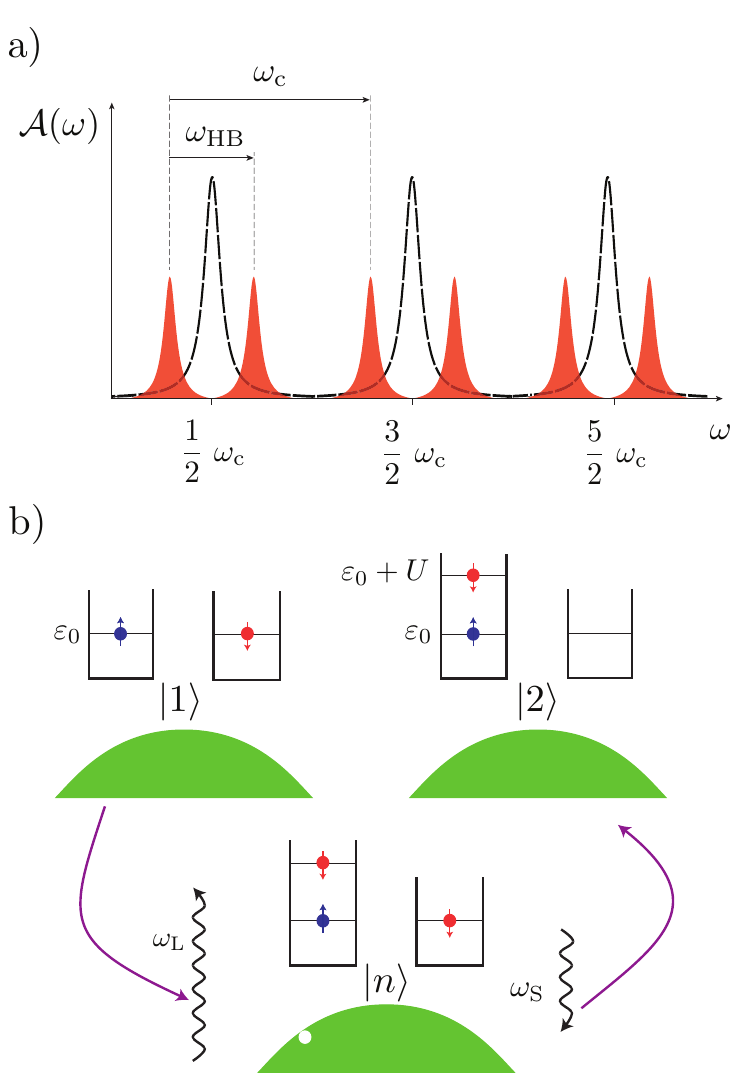}
\caption{a) A cartoon of the spectral function ${\cal A}(\omega)$ of the patterned/unpatterned 2DEG (red/black). 
	The Landau level peaks at $\omega = \omega_{\rm c}(n + 1/2)$ are split by on-site Coulomb interactions into Hubbard lower 
	and upper peaks, which are separated by $U \sim e^2/\ell_B$, where $\ell_B = \sqrt{\hbar c/(eB)}$ is the magnetic length.
	b) The relevant electronic process which contributes to the Raman scattering cross section. The initial state is labeled by $|1\rangle$, the final state by $|2\rangle$, while the intermediate state with {\it one hole} and an extra electron is labeled by $|n\rangle$. The final excited state is separated from the ground state by the Hubbard charge gap $U$, {\it i.e.} by the energy cost of having two antiparallel spin electrons on the same site. In the intermediate state we have also depicted the absorbed (at frequency $\omega_{\rm L}$) and emitted (at frequency $\omega_{\rm S}$) photons. The square wells denote two neighboring minima of the artificial-lattice potential. The core levels are not shown. The green areas denote valence-band electrons, which are assumed to be unaffected by the periodic modulation.  \label{fig:two}}
\end{figure}

\begin{figure*}[!ht]
	\centering
	\includegraphics[width=0.90\linewidth]{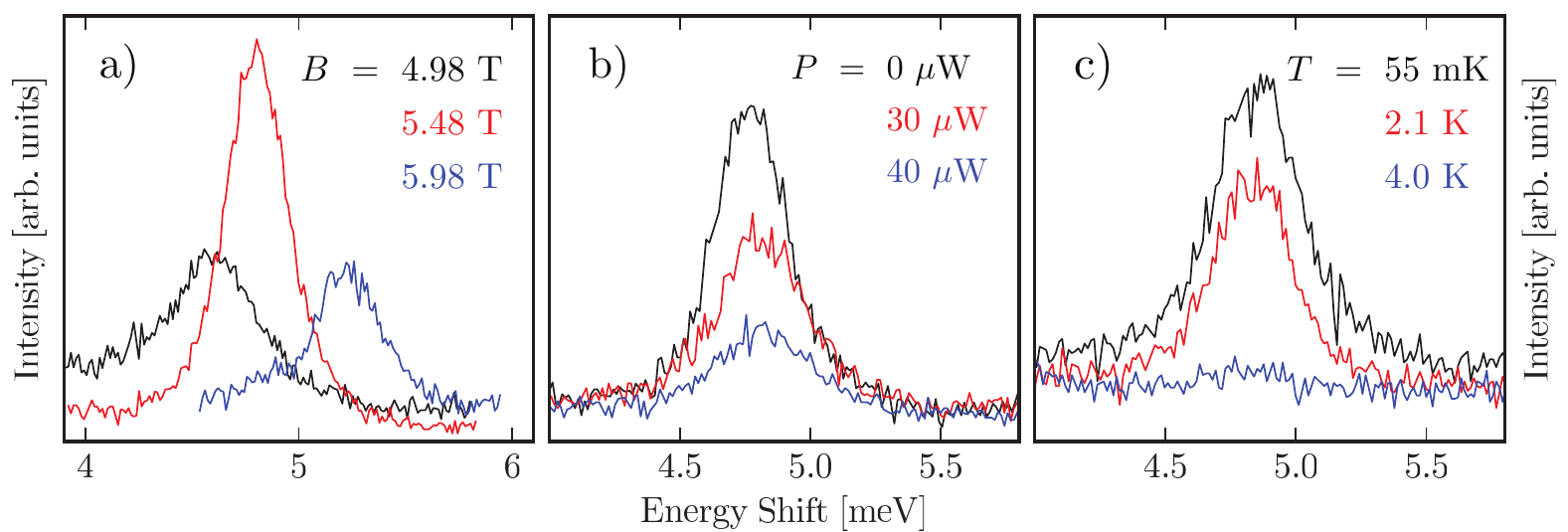}
	\caption{a) Resonant inelastic light scattering spectra of the Hubbard mode at three values of the magnetic field and $T = 1.7~{\rm K}$.
	b) Dependence of the Hubbard mode at $T = 1.5~{\rm K}$ 
	on the power $P$ (in $\mu {\rm W}$) of the HeNe laser used to photo-deplete the 
	two-dimensional electron gas. From black to blue the electron concentration per site decreases from $8 \pm 2$ to $3 \pm 2$ 
	(see \SM). Data in panel a) are at $P = 0~\mu {\rm W}$. 
	c) Temperature dependence of the Hubbard mode at $B = 5.48~{\rm T}$ and $P = 0~\mu {\rm W}$ 
	displaying an activated behavior with an activation energy of $0.2~{\rm meV}$.\label{fig:three}}
\end{figure*}

We identify the sublinear collective mode with a {\it Hubbard mode}, {\it i.e.} an excitation across split HBs. 
In the simplest scenario, this excitation emerges within the single-band Hubbard model~\cite{kotliar_rmp_2006,vollhardt_condmat_2010,hubbard_prsla_1964} that assumes a maximum concentration of two electrons per site. 
We proceed by first evaluating the Mott-Hubbard excitation gap as a function of $B$ and then we demonstrate that it weakly depends on electron concentration consistently with experimental data shown in Fig.~\ref{fig:three}b). Similar conclusions can be reached by employing multiband generalizations of the Hubbard model (see \SM).

The single-band Hubbard Hamiltonian encodes a competition between two energy scales: the kinetic energy $t$, which measures the overlap between electronic wavefunctions on neighboring lattice sites, and the interaction energy $U$, which measures the strength of the on-site Coulomb repulsion between two electrons:
\begin{equation}\label{eq:hamiltonian}
{\hat {\cal H}} = - t \sum_{\langle i,j\rangle} {\hat c}^\dagger_i {\hat c}_j  + \varepsilon_0 \sum_{i} {\hat n}_i + U \sum_i {\hat n}_{i\uparrow} {\hat n}_{i\downarrow}~.
\end{equation}
Here ${\hat c}^\dagger_i$ (${\hat c}_i$) creates (destroys) an electron at site $i$ (the sum in the first term is over all pairs of nearest-neighbor sites), and ${\hat n}_i = {\hat c}^\dagger_i {\hat c}_i$ is the local number operator; $\varepsilon_0$ denotes the energy of the single state which is available at each site $i$: this can be either empty, singly, or doubly occupied. In writing Eq.~(\ref{eq:hamiltonian}) we have neglected first-neighbor ({\it i.e.} inter-site) interactions. In the atomic, strongly correlated limit $U \gg t$, two split HBs emerge out of a single narrow band~\cite{hubbard_prsla_1964}. More precisely, this means that for $U \gg t$ the {\it spectral} function, {\it i.e.} the tunneling density-of-states, ${\cal A}(\omega)$ of the model described by Eq.~(\ref{eq:hamiltonian}) develops two peaks, one at $\hbar \omega =\varepsilon_0$ and one at $\hbar \omega  = \varepsilon_0 + U$. The emergence of HBs when the ratio $U/t$ increases from the weakly to the strongly correlated regime is accurately described by dynamical mean-field theory~\cite{kotliar_pt_2004,kotliar_rmp_2006,vollhardt_condmat_2010}.

In the experiments, the strongly correlated regime $U \gg t$ is achieved when $B$ quenches the hopping amplitude $t$ and increases the interaction energy $U$. The Hubbard-$U$ interaction scale can be written in terms of localized Wannier functions $\phi({\bm r})$ as
\begin{equation}\label{eq:UversusB}
U = \int d^2{\bm r} \int d^2{\bm r}' |\phi({\bm r})|^2V_{\rm ee}(|{\bm r} - {\bm r}'|)|\phi({\bm r}')|^2~,
\end{equation}
where $V_{\rm ee}(r) = e^2/(\epsilon r)$ is the long-range Coulomb interaction, with $\epsilon$ an effective dielectric constant. In the atomic limit the Wannier functions can be roughly approximated by a zero-angular-momentum wavefunction in the symmetric gauge, $\phi({\bm r}) = (2\pi \ell^2_B)^{-1/2}\exp{[-r^2/(4 \ell^2_B)]}$, where $\ell_B = \sqrt{\hbar c/eB}$ is the magnetic length. Simple algebraic manipulations on Eq.~(\ref{eq:UversusB}) yield
\begin{equation}\label{eq:UsqrtB}
U = \sqrt{\frac{\pi}{4}}~\frac{e^2}{\epsilon \ell_B}~,
\end{equation}
implying that, at least asymptotically, $U$ grows proportionally to $\sqrt{B}$. Microscopic details such as the precise shape of the confinement potential or the geometry of the lattice might affect the result (\ref{eq:UsqrtB}) quantitatively but not qualitatively: the scaling $\propto \sqrt{B}$ is robust in the asymptotic limit $\ell_B \ll 2r$, where $2r$ is the width of the potential minima of the artificial lattice [see Fig.~\ref{fig:one}a)].

The function ${\cal A}(\omega)$ in a $B$ field is pictorially illustrated in red in Fig.~\ref{fig:two}a). As a comparison, the black dashed line labels ${\cal A}(\omega)$ for an unpatterned 2DEG: we distinguish the usual Landau level peaks at frequencies $\omega_n = \omega_{\rm c}(n + 1/2)$ with integer $n$. In the nanopatterned sample these peaks are split into upper and lower Hubbard peaks by strong interactions. In this cartoon the measured cyclotron mode at $\omega_{\rm c} \propto B$ is an {\it inter}-Landau-level excitation. The measured sublinear mode seen in Fig.~\ref{fig:one}, instead, can be neatly explained as an {\it intra}-Landau-level excitation, which lies at a frequency $\omega_{\rm HB} = U/\hbar \propto \sqrt{B}$, between interaction-induced Hubbard peaks~\cite{katsnelson_JPCM_2010}.

Fitting the data labeled by red filled squares in Fig.~\ref{fig:one}d) with the simple functional form $\hbar\omega_{\rm HB} = \alpha \sqrt{B[{\rm T}]}$ we find $\alpha \sim 2~{\rm meV}$, thereby providing a direct measurement of the Hubbard-$U$ on-site energy scale for our nanopatterned 2DEG. The measured $U$ is a factor of two smaller than the value extracted from Eq.~(\ref{eq:UsqrtB}) with the high-frequency GaAs dielectric constant $\epsilon = 13$.  In Fig.~\ref{fig:two}b) we illustrate a possible two-photon process that contributes to the scattering cross section of the HB collective mode. The calculated scattering cross section decays exponentially for sufficiently large values of $B$ (see \SM), in agreement with the data reported in Fig.~\ref{fig:three}a). In contrast to the cyclotron mode, indeed, the intensity of the Hubbard mode increases up to $B \simeq 5.5~{\rm T}$ and then collapses exponentially at larger fields.

The Hubbard mode energy exhibits a rather weak dependence on electron concentration, see Fig.~\ref{fig:three}b), which is decreased by the photo-depletion technique (see \SM). In the atomic limit the dependence of the Mott-Hubbard gap $\hbar\omega_{\rm HB}$ on electron concentration is indeed a small effect, of the first order in the parameter $t/U$. In the limit of vanishing electron concentration, the strength of the transition between the two HBs also vanishes because there are no available states in the upper HB (see \SM). This is in agreement with the large dependence of the intensity of the Hubbard mode on electron density reported in Fig.~\ref{fig:three}b). Finally, the Hubbard mode displays a large sensitivity to temperature changes [see Fig.~\ref{fig:three}c)] and disappears close to $T = 5~{\rm K}$.

\begin{figure}[!ht]
\centering
\includegraphics[width=0.90\linewidth]{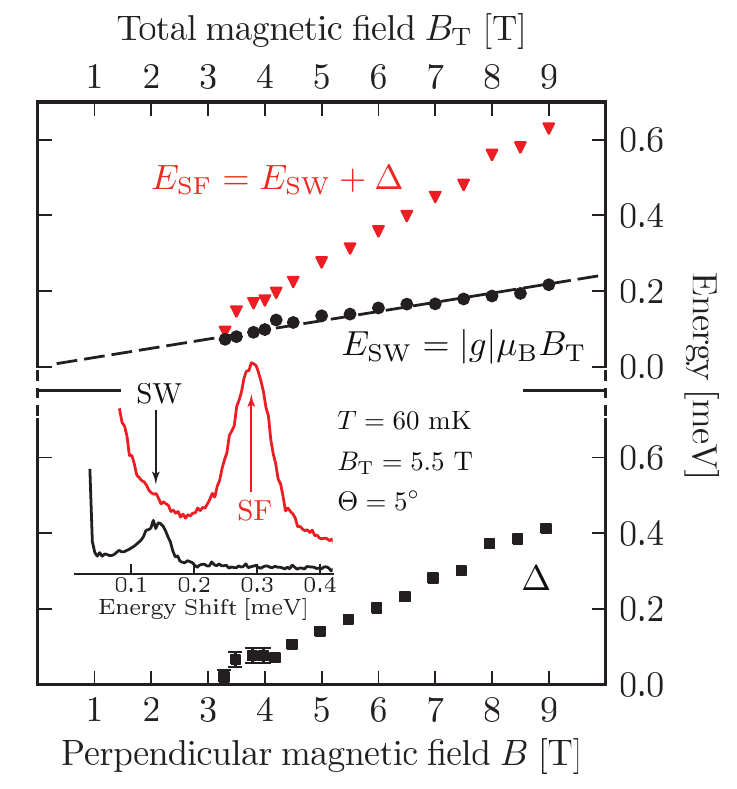}
\caption{Energies of the spin-wave mode (black filled circles) and the higher energy spin-flip mode (red filled triangles). 
The black dashed line is a linear fit to the data using the standard Zeeman formula: $E_{\rm SW} = |g| \mu_{\rm B} B_{\rm T}$. We find $|g| = 0.42$, in agreement with the value expected for GaAs. 
Representative examples of the two spin excitation modes at two different laser energies
(red line: $1522.6~{\rm meV}$; black line: $1522.4~{\rm meV}$) are reported in the inset.
The black filled squares label the splitting $\Delta $ between the two spin modes. \label{fig:four}}
\end{figure}

We now focus on the low-energy portion of the excitation spectra, {\it i.e.} $\hbar \omega < 1~{\rm meV}$. In ordinary 2DEGs this sector is characterized by the spin-wave (SW) mode, a spin-flip excitation across the spin gap that, at long wavelength, occurs at the bare Zeeman energy $g\mu_{\rm B} B_{\rm T}$, where $\mu_{\rm B}$ is the Bohr magneton, $g$ is the Land\'e gyromagnetic factor, and $B_{\rm T} = \sqrt{B^2_\| + B^2}$ is the total magnetic field [see Fig.~\ref{fig:one}b)]. The inset to Fig.~\ref{fig:four} shows a representative result at $B_{\rm T} = 5.5~{\rm T}$. ÊThe SW mode is visible at energies close to 0.15 meV. The SW energy versus total field is reported in Fig.~\ref{fig:four} as black filled circles. The spin mode is not visible below $B_{\rm T} = 3~{\rm T}$.

The inset to Fig.~\ref{fig:four} displays an additional strong and sharp mode {\it above} the SW, which has no counterpart in an unpatterned 2DEG. The energy dependence of this mode is shown in Fig.~\ref{fig:four} as red filled triangles. The splitting $\Delta$ of this mode from the SW (filled squares) occurs above a threshold $B$ value and depends on the perpendicular magnetic field {\it only}, a fact that underlines the pivotal role of electron-electron interactions. The two modes disappear at temperatures approaching $T = 1~{\rm K}$.

The observation of a spin doublet is very intriguing and suggests the occurrence of a novel correlated state with a gap $\Delta$. 
Different types of Coulomb-driven broken-symmetry scenarios have been proposed in the context of graphene at large magnetic fields~\cite{nomura_prl_2006,alicea_prb_2006,gusynin_prb_2006,abanin_prl_2006,chakraborty_review_2010} and linked to observations of gap openings in magneto-transport experiments~\cite{zhang_prl_2006,checkelsky_prl_2008,giesbers_prb_2009}. Remarkably, one of these scenarios~\cite{alicea_prb_2006} predicts a splitting of the SW mode similar to what we see in our experiment~\cite{scalinginB} associated to the occurrence of lattice-scale order in the honeycomb lattice.

In the case of graphene, however, the high-field regime is not experimentally accessible: one is always in the weak-field regime, that is, the magnetic length is much larger than the interatomic distance. This is not the case in our artificial honeycomb lattice. 
To support the existence of graphene-like effects in our system, we carried out 
calculations of the density-of-states based on a tight-binding model~\cite{yuan_prb_2010} in the presence of disorder comparable 
to the hopping energy and in the ultra high-magnetic-field regime ($\ell_B < a$).  These results reveal the persistence of a structure reminiscent of the zero-energy Landau level of graphene (see \SM). Similarly to what happens in graphene~\cite{zhang_prl_2006,checkelsky_prl_2008,giesbers_prb_2009,chakraborty_review_2010}, electron-electron interactions 
can lead to a re-organization of this low-energy sector yielding a broken-symmetry ground state with an energy gap $\approx \Delta$.
Additionally, the observed opening of the gap above a threshold magnetic field indicates a delicate interplay between hopping, disorder and many-body effects.

\acknowledgments

We acknowledge financial support by the Project ``Knowledge Innovation Program" (PKIP)
of the Chinese Academy of Sciences, Grant No. KJCX2.YW.W10 (M.P. and G.V.),  
FOM (the Netherlands) (S.Y. and M.I.K.), 
the National Science Foundation (NSF) grants DMR-0803691 and CHE-0641523 (A.P.), and the Italian Ministry of research through the FIRB and COFIN programs (V.P.). 
S.Y. and M.I.K. acknowledge computer time from NCF (the Netherlands). We wish to thank Rosario Fazio, Allan MacDonald, 
and Pasqualantonio Pingue for useful conversations.

\begin{appendix}
\section{SUPPLEMENTARY MATERIAL}\label{SM}
This section contains technical details and numerical results relevant to the main text.

\section {\bf Nanofabrication of the artificial lattice and resonant enhancement of light scattering peaks}

The in-plane potential modulation is achieved by defining an array of Nickel disks (with diameter $2r$) arranged in a honeycomb-lattice geometry (with lattice constant $a$) by e-beam nanolithography and then by etching away the material outside the disks by inductive coupled reactive ion shallow etching~\cite{garcia,gibertini_prbr_2009}. Owing to the dependence of band-bending profiles on GaAs cap-layer thickness, the resulting pillars [see Fig.~\ref{fig:one}a) in the main text] induce a lateral potential modulation with an amplitude of a few ${\rm meV}$'s acting on the electronic system~\cite{gibertini_prbr_2009}. By tuning the etching depth $d$ we can reach different regimes ({\it i.e.} different values of $V_0$). We focus on a sample with $d \sim 60~{\rm nm}$ for which we estimate $V_0 \sim 4~{\rm meV}$~\cite{DeSimoni2010}.

The resonant inelastic light scattering experiments were performed by using a ring-etalon Ti:Sapphire laser with a 
tunable wavelength of circa $800~{\rm nm}$ in resonance with the magneto-luminescence of the nanostructured semiconductor (data not shown) focused on the array with a 100 $\mu$m diameter area. The scattered light was collected into a triple grating
spectrometer with CCD detection: for the geometry of the experiment see Fig.~\ref{fig:one}b) in the main text.
The intensity of the incident radiation was kept to values well below $10^{-1}~{\rm W}/{\rm cm}^{2}$ to avoid significant heating of the electrons.

\begin{figure}[!t]
\centering
\includegraphics[width=0.90\linewidth]{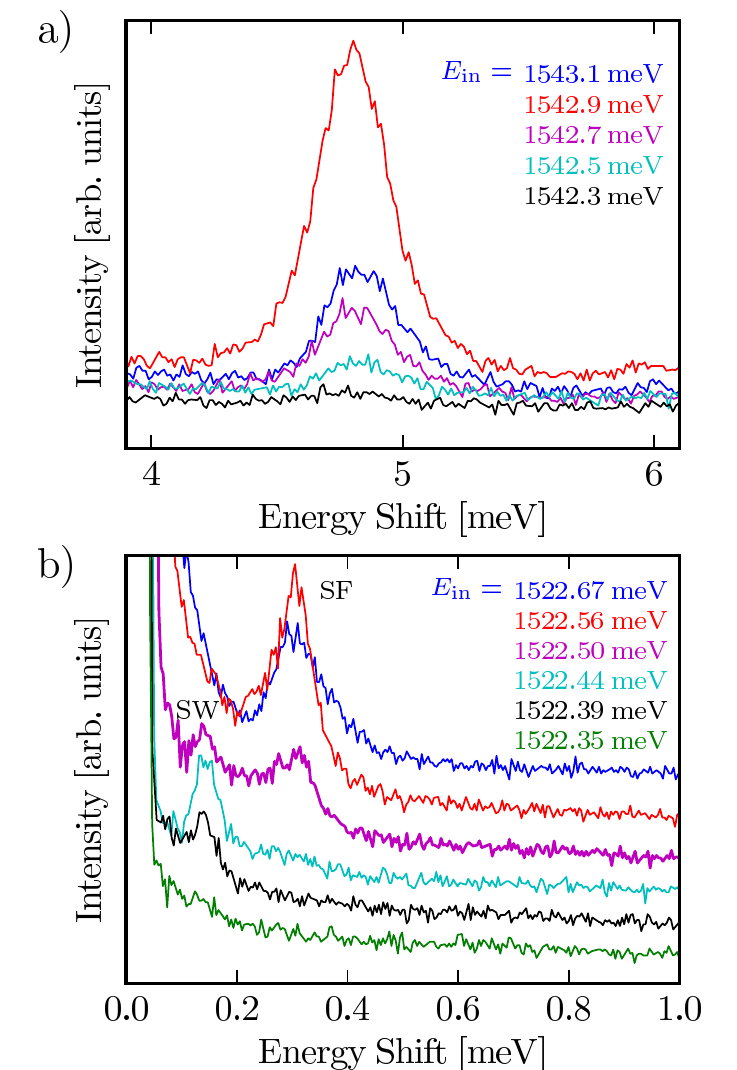}
\caption{Resonant profile of inelastic light scattering spectra showing the Hubbard mode at $T = 1.7~{\rm K}$ and $B = 5.48~{\rm T}$ [(panel a)] and the spin-wave (SW) and spin-flip (SF) modes at $T = 55~{\rm mK}$ and $B= 5.48~{\rm T}$ [panel b)]. The incident photon energies are indicated.\label{fig:onesuppl}}
\end{figure}

Crucial for the optical detection of the Hubbard and spin modes of the electrons in the honeycomb lattice is the 
resonant enhancement of the corresponding light scattering cross sections. This occurs when the 
incoming laser energy matches the energy of an inter-band transition of the host crystal. In the experiment this is achieved by tuning the laser energy and plotting the light scattering signal as a function of energy shift with respect to the laser energy. Examples of the resonant effect for the Hubbard mode and for the spin-wave and spin-flip modes (reported in Figs.~1, 3, and~4 of the main text) can be found in Fig.~\ref{fig:onesuppl}. When displayed as a function of energy shift, the collective modes occur at a fixed energy and their intensities follow a resonant profile.
\section{\bf Light scattering spectroscopy of the cyclotron mode}

\begin{figure}[!t]
\centering
\includegraphics[width=0.90\linewidth]{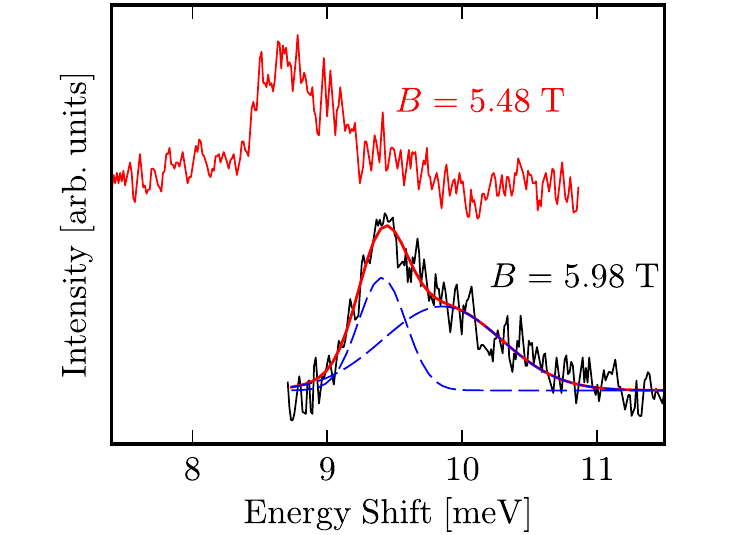}
\caption{Representative resonant inelastic light scattering spectra showing the cyclotron mode at $T = 1.7~{\rm K}$ and $B = 5.48~{\rm T}$ and $5.98~{\rm T}$. Incident photon energy is $1547.7~{\rm meV}$ and $1546.5~{\rm meV}$, respectively. A fit (red solid line) with two gaussians (blue dashed lines) is also shown.\label{fig:twosuppl}}
\end{figure}

Representative resonant inelastic light scattering spectra at $T = 1.7~{\rm K}$ manifesting the cyclotron mode at two values of the perpendicular magnetic field $B$ are reported in Fig.~\ref{fig:twosuppl}. The cyclotron mode is parity-forbidden in the dipole approximation, and its non-zero albeit weak cross section originates from valence-band mixing effects~\cite{pinczuk92}. The fact that we see the {\it ordinary} 2DEG cyclotron mode in our nanopatterned system is due to a magnetic length $\ell_B$ that is smaller than the lattice parameters in the explored range of magnetic fields. (The magnetic length $\ell_B$ at $B=2~{\rm T}$ is $\sim 18~{\rm nm}$, which is smaller than the pillar diameter $2r \sim 60~{\rm nm}$.) There is a {\it second} peak above $\omega_{\rm c}$ (the red solid line in Fig.~\ref{fig:twosuppl} is a fit with two gaussians, which are shown as blue dashed lines) that can be associated to a finite-wavevector excitation due to the dispersion of the cyclotron mode~\cite{pinczuk92} and the optical grating effect of the lattice~\cite{sohn}. The cyclotron mode can be detected for $B\gtrsim 2~{\rm T}$, a value associated with electrons occupying the lowest Landau level only as indicated by the magneto-transport analysis (data not shown and Ref.~\onlinecite{DeSimoni2010}). We found that its intensity does not vary significantly in the explored range of magnetic fields.

\section {\bf Evaluation of the scattering cross section of the Hubbard mode}

In Fig.~\ref{fig:two}b) of the main text we have shown a possible two-photon process that contributes to the scattering cross section of the Hubbard collective mode. In this scheme the initial state $|1\rangle$ is characterized by a pair of neighboring singly-occupied sites with two active electrons in an antiparallel spin configuration. In the final state $|2\rangle$ the two initial electrons occupy the same site. The final excited state is separated from the ground state by the Hubbard
charge gap $U$, {\it i.e.} by the energy cost of having two antiparallel spin electrons on the same site. Because the lower and upper states split by $U$ have the same parity (as they both emerge from the very same single-particle level at energy $\varepsilon_0$) the process depicted in Fig.~2B of the main text is allowed in the dipole approximation, and its intensity is thus expected to be much larger than the one of the cyclotron mode at $\omega_{\rm c}$ [because of the vanishing of the dynamical structure factor ${\cal S}(\omega)$ at $\omega_{\rm c}$] in agreement with the experimental observation. The scattering cross section can be roughly evaluated in the dipole approximation~\cite{book:landau_QED} by assuming that the valence-band holes are not strongly affected by the external periodic modulation. Within the same degree of accuracy used to derive Eq.~(3) in the main text, we find that the cross section decays exponentially, {\it i.e.} $d\sigma/d\Omega' \propto \exp{[-a^2/(4 \ell^2_B)]}$, for sufficiently large values of $B$ in agreement with the experimental observation.

\section {\bf Reducing the electron density by the photo-depletion technique}

The electron density of a modulation-doped quantum well like the one used in our study can be reduced by continuous illumination 
with photon energy larger than the quantum-well (QW) barrier ($\approx 1.6~{\rm eV}$ in our case). 
The mechanism of photo-depletion is based on the fact that photoexcited electrons in the AlGaAs barrier contribute to a charge compensation of the ionized donors while photoexcited holes are swept in the QW region, thereby reducing the electron density through electron-hole radiative recombination~\cite{photodepletion}. 

\begin{figure}[!t]
\centering
\includegraphics[width=0.90\linewidth]{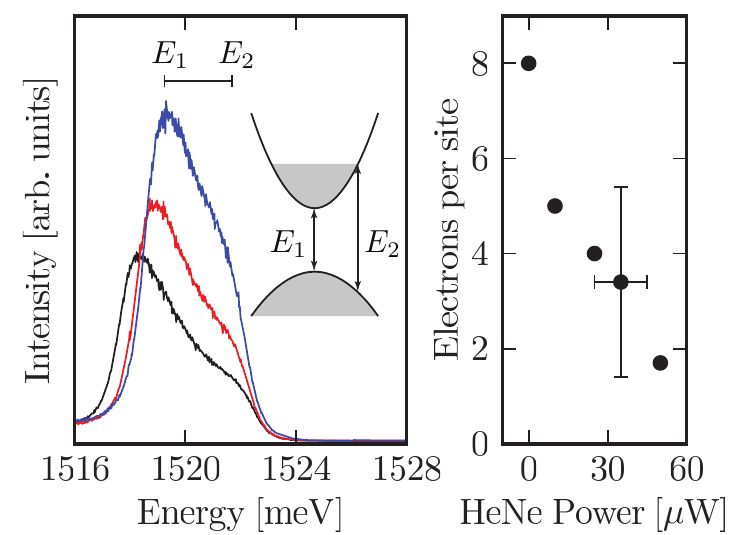}
\caption{(Left panel) Evolution of the photoluminescence spectra of the unpatterned 2DEG with varying HeNe power $P$ (black $\to 0~\mu {\rm W}$; red $\to 10~\mu {\rm W}$; blue $\to 25~\mu {\rm W}$). The inset shows a schematic profile of the conduction and valence bands: occupied states are indicated by gray-shaded areas. $E_1$ and $E_2$ refer to the main structures in the photoluminescence. (Right panel) Estimated electron concentration per site in the patterned region of the sample {\it versus} HeNe laser power.\label{fig:threesuppl}}
\end{figure}

In our studies we used a HeNe laser at energy of $1.96~{\rm eV}$ with powers $P$ up to $50~\mu {\rm W}$. The calibration of electron density versus HeNe power $P$ is obtained by monitoring the evolution of the QW photoluminescence (PL) in an unpatterned region of the sample. Representative spectra are shown in Fig.~\ref{fig:threesuppl}.  The QW optical emission exhibits the usual line shape of a modulation-doped QW luminescence, as determined by the recombination of electrons from the bottom of the subband up to the Fermi energy (see inset to the left panel of Fig.~\ref{fig:threesuppl}). The distance between the main PL peak at energy $E_1$ and the shoulder at energy $E_2$ is related to the electron density $n$ through the relation $|E_1 - E_2| = E_{\rm F}(1+ m_{\rm e}/m_{\rm h})$, where $m_{{\rm e}({\rm h})}$ are the effective masses for electrons (holes) and $E_{\rm F} = \pi n\hbar^2/m_{\rm e}$ is the Fermi energy. 
This relation thus allows us to estimate the electron density $n$ in the unpatterned 2DEG as a function of $P$. This simple analysis cannot be applied to the patterned region since in this case the PL lineshape is significantly modified as discussed in Ref.~\onlinecite{gibertini_prbr_2009}.
The 2DEG density $n$ can be significantly reduced from its zero-power value ($\sim 1 \times 10^{11}~{\rm cm}^{-2}$) as the HeNe power $P$ is increased. 

The right panel in Fig.~\ref{fig:threesuppl} shows the estimated electron concentration per site at different HeNe powers in the nanopatterned region using the calibration method described above (but starting from the $P = 0$ value of the electron density $\sim 3-4 \times 10^{10}~{\rm cm}^{-2}$ of the nanopatterned sample).

\begin{figure}[!t]
\centering
\includegraphics[width=0.90\linewidth]{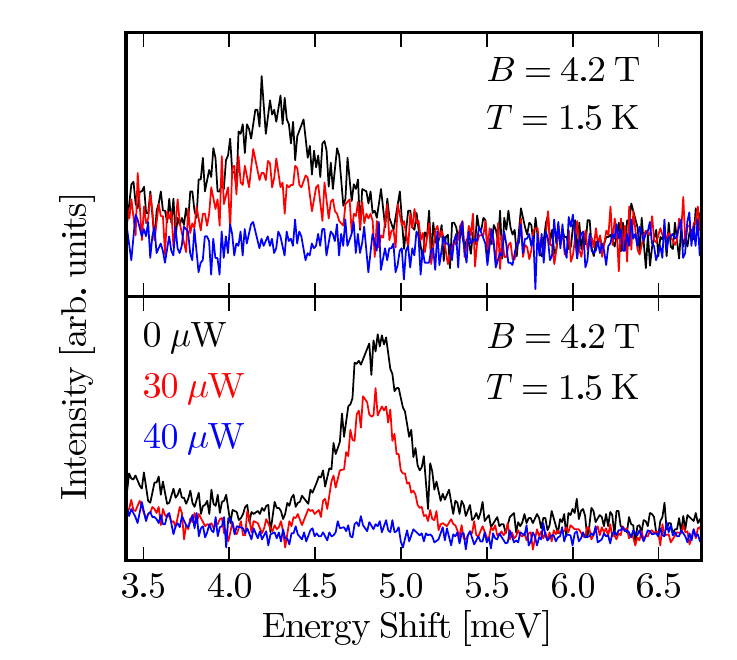}
\caption{Representative inelastic light scattering data showing the evolution of the Hubbard mode versus HeNe power 
at two different magnetic fields. The density is decreased (from black to blue) by means of the photo-depletion technique.\label{fig:foursuppl}}
\end{figure}
\section {\bf Evaluation of density dependence of the Hubbard mode}

In this Section we demonstrate that the Mott-Hubbard gap $\hbar \omega_{\rm HB}$ 
depends very weakly on the electron concentration in the single-band Hubbard model and in the atomic limit ($U\gg t$).

In the so-called ``Hubbard I" approximation~\cite{hubbard_classic}, 
which works well at strong coupling, the Hubbard bands $E_{\pm}({\bm k})$ for spin $\sigma$ electrons 
are given by the following expression:
\begin{eqnarray}\label{eq:Hubbard1bands}
E_{\pm}({\bm k}) &=& \varepsilon_0 + \frac{1}{2}\Big[ t({\bm k}) + U \nonumber\\ 
&\pm& \sqrt{t^{2}({\bm k})+U^{2}+2t({\bm k})U(2n_{-\sigma} -1)}\Big]
\end{eqnarray}
where $t({\bm k})$ is the bare band energy and $n_{-\sigma}$ is the concentration of electrons with spin projection opposite to $\sigma$. In the atomic limit Eq.~(\ref{eq:Hubbard1bands}) simplifies to
\begin{equation}
\left\{
\begin{array}{l}
{\displaystyle E_{+}({\bm k}) = \varepsilon_0 + U\left[1 + \frac{t({\bm k})}{U} n_{-\sigma}\right]}\vspace{0.2 cm}\\
{\displaystyle E_{-}({\bm k}) = \varepsilon_0 + t({\bm k})(1-n_{-\sigma})}
\end{array}
\right.~.
\end{equation}
We thus immediately see that one band is centered at $\varepsilon_0 + U$ and the other one at $\varepsilon_0$, 
independently of the electron concentration. The dependence of the Mott-Hubbard gap $\hbar \omega_{\rm HB} \equiv {\rm min}_{\bm k} [E_+({\bm k}) - E_-({\bm k})] = U + {\cal O}(t/U)$ on electron concentration is a small effect, of the first order in $t/U$.
In the limit $n_{-\sigma} = 0$ the two Hubbard band energies are at $\varepsilon_0 +t({\bm k})$ and $\varepsilon_0 + U$, but the upper band has a vanishing number of states ($n_{-\sigma}$) associated with it. In other words, the strength of the transition from the lower to the upper band vanishes because there are no available states in the upper band.

A similar result is valid also within the multiband generalization of the ``Hubbard I" approximation~\cite{hubbard_subsequent}. 
The Green's function in the strongly-correlated limit is determined by the equation
\begin{equation}
G^{-1}(E, {\bm k}) = G^{-1}_{\rm at}(E) - t({\bm k})~,
\end{equation}
where $G_{\rm at}(E)$ is the energy-dependent Green's function of the atomic problem. One can prove (see the Appendix in the work by Lebegue {\it et al.}~\cite{hubbard_subsequent}) that this expression is exact for the multiband Hubbard model up to first order in $t/U$. 
The poles of this function, which determine the positions of the centers of the Hubbard bands, do not depend on electron concentration, while the residues, which determine the width of the Hubbard bands, do. 

These theoretical arguments are fully consistent with the experimental observations shown in Fig.~\ref{fig:three}b) of the main text and in Fig.~\ref{fig:foursuppl}.

\section {\bf Tight-binding calculations in the ultra-high-magnetic-field regime} 

We have performed extensive tight-binding calculations of the density-of-states (DOS) of non-interacting electrons hopping on a honeycomb lattice in the presence of a perpendicular magnetic field $B$ and of uniformly-distributed disorder of amplitude $W$. We have simulated both on-site disorder, which is mathematically described by an Hamiltonian of the type
\begin{equation}\label{eq:onsite}
{\hat {\cal H}}_{\rm dis} = \sum_i \varepsilon_i {\hat n}_i
\end{equation}
with $\varepsilon_i$ uniformly distributed in the interval $[-W/2, +W/2]$ and disorder due to random hopping, which is described by an Hamiltonian of the type
\begin{equation}\label{eq:randomhopping}
{\hat {\cal H}}'_{\rm dis} = \sum_{\langle i,j\rangle} \delta t_{ij} {\hat c}^\dagger_i c_j
\end{equation}
where $\delta t_{ij}$ is uniformly distributed in the interval $[-W/2, +W/2]$. Details on the numerical technique can be found {\it e.g.} in Ref.~\onlinecite{yuan_prb_2010}. 

\begin{figure}[!t]
\centering
\includegraphics[width=0.90\linewidth]{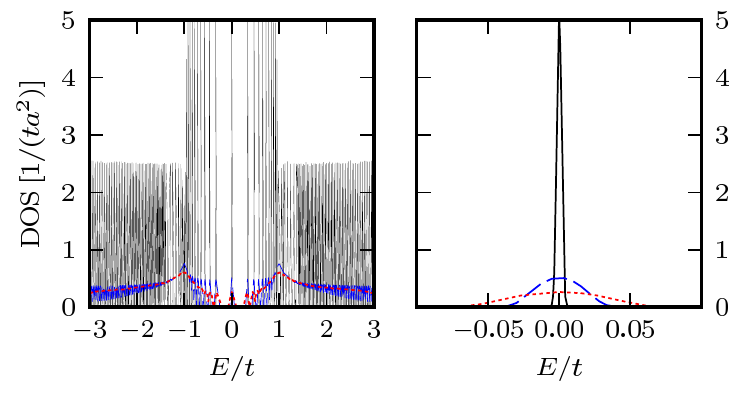}
\caption{Calculated DOS [in units of $1/(ta^2)$] as a function of energy $E$ (in units of $t$) for electrons hopping on a honeycomb lattice with lattice constant $a = 130~{\rm nm}$ and in the presence of a small magnetic field $B = 0.001~{\rm T}$. We clearly see the zero-energy Landau level, unevenly spaced Landau levels of massless Dirac fermions at low energies, and equally spaced ordinary Landau levels at higher energies. The panel on the right is a zoom of the low-energy sector. Black lines label the DOS of the clean system, while blue dashed ($W =0.5~t$) and red dotted ($W=1.0~t$) lines label results for the disordered system. Data for $W \neq 0$ have been obtained by using random on-site disorder, see Eq.~(\ref{eq:onsite}). These numerical results have been obtained by using periodic boundary conditions on a sample containing $3200 \times 3200$ lattice sites. \label{fig:fivesuppl}}
\end{figure}

\begin{figure}[!t]
\centering
\includegraphics[width=0.90\linewidth]{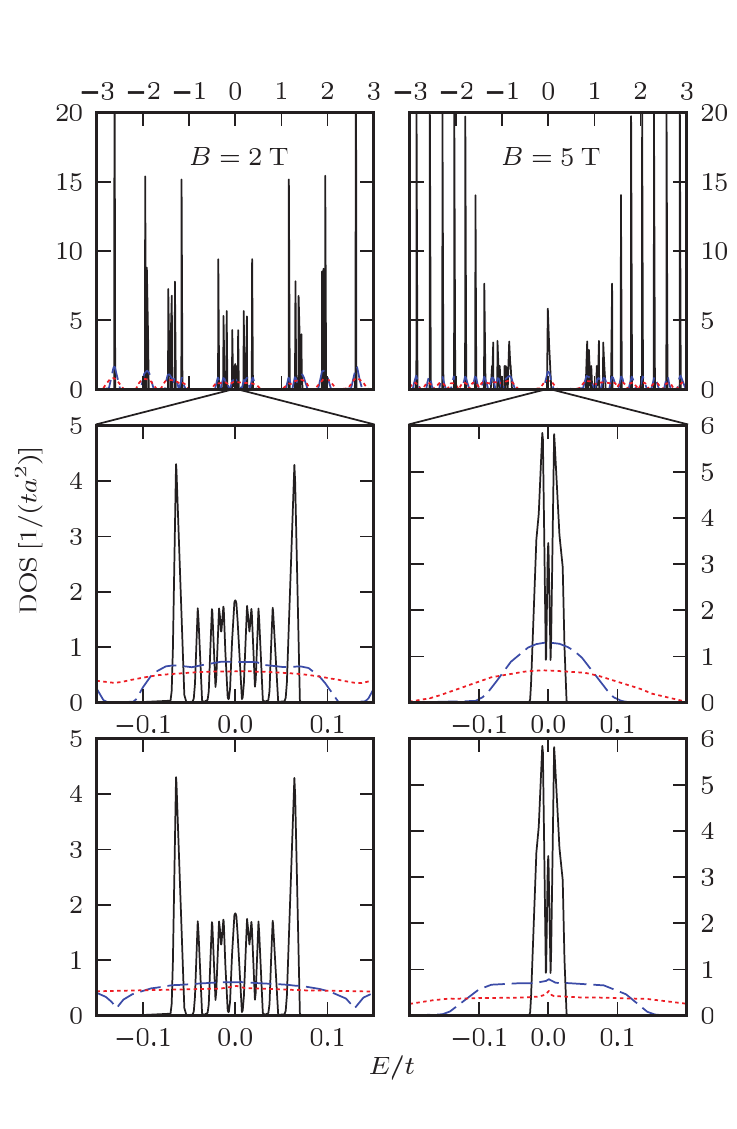}
\caption{Same as in Fig.~\ref{fig:fivesuppl} but for higher values of the magnetic field, $B=2~{\rm T}$ (left panels) and $B=5~{\rm T}$ (right panels). Color coding and labeling are the same as in Fig.~\ref{fig:fivesuppl}. Data for $W \neq 0$ in the two panels at the bottom have been obtained by using disorder due to random hopping, see Eq.~(\ref{eq:randomhopping}).\label{fig:sixsuppl}}
\end{figure}

In Figs.~\ref{fig:fivesuppl}-\ref{fig:sixsuppl} we collect our main findings for a honeycomb lattice with $a = 130~{\rm nm}$ and $B$ varying from $0.001~{\rm T}$ up to $5~{\rm T}$. 

In Fig.~\ref{fig:fivesuppl} we clearly see that when $\ell_B > a$ the calculated DOS is indistinguishable from that of graphene and exhibits a zero-energy Landau level and massless Dirac fermions at low energies stemming from the topology of the lattice~\cite{mcclure_pr_1956,katsnelson_ssc_2007}. When disorder is switched on, the zero-energy Landau level broadens. 
Fig.~\ref{fig:fivesuppl} illustrates the DOS in the ultra-high-magnetic-field regime, {\it i.e.} for $\ell_B < a$. For $B= 2~{\rm T}$ and $5~{\rm T}$, for example, we clearly see that, in the absence of disorder, a series of Hofstadter states appears near zero energy due to commensurability effects~\cite{claro_prb_1979,macdonald_prb_1984,rammal_jphys_1985} ({\it i.e.} the magnetic flux $\Phi = 3\sqrt{3} a^2 B/2$ through the unit cell of the honeycomb lattice being of the same order of the quantum flux unit $hc/e$). When disorder is taken into account a broad structure near zero energy emerges akin to the zero-energy Landau level in disordered graphene (see right panel in Fig.~\ref{fig:fivesuppl}). Notice that this structure has a width in energy of the order of $0.2~t$ for $W = 0.5~t$, {\it i.e.} roughly twice the width of the zero-energy Landau level in graphene for the same value of $W$.

In analogy to what observed experimentally in graphene~\cite{zhang_prl_2006,checkelsky_prl_2008,giesbers_prb_2009,chakraborty_review_2010}, we expect that electron-electron interactions can strongly re-organize this low-energy sector yielding a correlated ground state with an energy gap of the order of the spin-doublet energy splitting $\Delta$ introduced in the main text. Similarly to graphene~\cite{chakraborty_review_2010}, the precise mechanism responsible for the emergence of this novel ground state is, to date, unknown. A broken-symmetry ground state driven by lattice-scale interactions beyond the continuum model is compatible with the experimental results~\cite{alicea_prb_2006}.

Before concluding, we would like to emphasize that in the limit in which $\ell_B$ is smaller than $a$ effects beyond those captured by the 
tight-binding model should be taken into account since the shape of the ``atomic orbitals" is altered. 
To a first approximation, the magnetic field simply increases the confinement of electrons in each minimum of the periodic
potential (``magnetic squeezing") thereby renormalizing the hopping amplitude $|t| \to |t(B)|$. In the clean
limit this fact has no implications on the validity of our results in Figs.~\ref{fig:fivesuppl}-\ref{fig:sixsuppl}. 
Indeed, the data reported in these plots illustrate the DOS [in units of $1/(t a^2)$] as a function of $E/t$: 
the hopping amplitude is the only energy scale and its precise value thus does not matter.

In the presence of disorder, however, as magnetic field $B$ increases, the tendency to Anderson
localization increases too. The actual phase diagram (and broken-symmetry states) of the nanopatterned electron gas 
results from the competition between electron-electron
interactions, disorder, and external periodic potential. This is clearly beyond the scope of the elemental 
theoretical analysis reported in the main text and in the \SM and is left for future investigations. 
The experimental data in Fig.~\ref{fig:four} of the original
manuscript strongly suggest that electron-electron interactions are capable of re-organizing the low-energy
degrees of freedom of the nanopatterned electron liquid at least up to magnetic field values of the order of 8-9 Tesla.
\end{appendix}

\end{document}